\documentclass{appolb}

\usepackage{graphicx}
\usepackage{amsmath,amssymb}
\usepackage{amsfonts}
\usepackage{physics}
\usepackage{xspace}
\usepackage{bm}
\usepackage{mathrsfs}
\usepackage{nicefrac}

\usepackage[dvipsnames]{xcolor}
\usepackage{multirow}
\usepackage[format=plain,justification=RaggedRight,labelsep=endash,font=small,labelfont=sc]{caption}
\newcommand{\pdagger}{{\phantom{\dagger}}}

\usepackage[T1]{fontenc}

\usepackage{graphicx}
\usepackage{flushend}
\usepackage[numbers,sort&compress]{natbib}
\usepackage[colorlinks,citecolor=blue,urlcolor=blue,linkcolor=blue]{hyperref}

\begin{document}

\title{Quantum baryon number fluctuations in subsystems of a hot and dense relativistic gas of fermions%
%\thanks{Presented at ...}%
}
\author{Arpan Das
\address{Institute of Nuclear Physics Polish Academy of Sciences, PL-31-342 Krakow, Poland}\\
Wojciech Florkowski
\address{Institute of Theoretical Physics, Jagiellonian University, PL-30-348 Krakow, Poland}\\
Radoslaw Ryblewski
\address{Institute of Nuclear Physics Polish Academy of Sciences, PL-31-342 Krakow, Poland}\\
Rajeev Singh
\address{Institute of Nuclear Physics Polish Academy of Sciences, PL-31-342 Krakow, Poland}}
%\\
%{Third Author of different affiliation
%}
%the Name(s) of other Author(s)
%\address{affiliation}
%}

%\title{Insert your title here\thanksref{t1}}
%\title{Quantum fluctuations of energy in subsystems of a hot relativistic gas}


%\author{Arpan Das\thanksref{e1,addr1}
%        \and
%        Wojciech Florkowski\thanksref{e2,addr2}
%        \and
%        Radoslaw Ryblewski\thanksref{e3,addr1}
%        \and
%        Rajeev Singh\thanksref{e4,addr1}%etc.
%}

%\thankstext[$\star$]{t1}{Thanks to the title}
%\thankstext{e1}{e-mail: arpan.das@ifj.edu.pl}
%\thankstext{e2}{e-mail: wojciech.florkowski@uj.edu.pl}
%\thankstext{e3}{e-mail: radoslaw.ryblewski@ifj.edu.pl}
%\thankstext{e4}{e-mail: rajeev.singh@ifj.edu.pl}

%\institute{Institute of Nuclear Physics Polish Academy of Sciences, PL-31-342 Krakow, Poland\label{addr1}
%          \and
%          Institute of Theoretical Physics, Jagiellonian University, PL-30-348 Krakow, Poland\label{addr2}
%}

\date{Received: date / Accepted: date}

\maketitle

\begin{abstract}
Quantum features of the baryon number fluctuations in subsystems of a hot and dense relativistic gas of fermions are analyzed. We find that the fluctuations in small systems are significantly different compared to their values known from the statistical physics, and diverge in the limit where the system size goes to zero. The numerical results obtained for a broad range of the thermodynamic parameters expected in heavy-ion collisions are presented. They can be helpful to interpret and shed new light on the experimental data. 
\end{abstract}

\section{Introduction}
\label{sec:intro}
%%%%%%%%%%%%%%%%%%%

Statistical fluctuations in many-body systems (microscopic as well as macroscopic ones) play a very important role in physics as they encode the crucial information about possible phase transitions, dissipation, and clustering phenomena~\cite{Smoluchowski,PhysRevLett.85.2076,Huang:1987asp,Kubo1,Lifshitz:1963ps,PhysRevLett.49.1110}. An unexplored novel feature of the fluctuations is their increase for small systems in the cases where the quantum effects become important. Such effects have been analyzed quantitatively in our two recent papers~\cite{Das:2020ddr,Das:2021aar} where we addressed the fluctuations of the energy density in a hot gas of bosons and fermions. Our results indicate limitations of the concept of the fluid elements used in relativistic hydrodynamics applied in the description of heavy-ion collisions. As the size of a subsystem drops below about 0.5 fm, the energy density fluctuations (for typical values of temperatures and particle masses) become so large that they are comparable with their mean values. In this case, the physical picture of fluid cells with well defined energy density becomes unjustified. An interesting feature of our calculations~\cite{Das:2020ddr,Das:2021aar} is that not only fluctuations diverge for the subsystem with size approaching zero, but also quantum statistical fluctuations agree with thermodynamic fluctuations if the subsystems become sufficiently large~\cite{Das:2020ddr,Das:2021aar}. In this way, our approach incorporates quantum features into statistical mechanics in a very natural way.

In the present work, following similar ideas to that developed in~\cite{Das:2020ddr,Das:2021aar}, we discuss fluctuations of the baryon number density in a hot and dense relativistic gas of fermions. Our analysis is relevant for relativistic heavy-ion physics, in particular, in the context of the beam energy scan (BES). Hunt for the conjectured critical endpoint in the QCD phase diagram has triggered vast theoretical and experimental studies of many fluctuation observables. For example, one studies fluctuations of the conserved quantum numbers in QCD, such as the baryon number, electric charge, or strangeness. They all provide an excellent opportunity to study the critical phenomena~\cite{Berges:1998rc,Halasz:1998qr,Stephanov:1998dy,Stephanov:1999zu,Hatta:2002sj,Stephanov:2008qz,Berdnikov:1999ph,Kitazawa:2013bta}. 

Following our previous framework~\cite{Das:2020ddr,Das:2021aar}, we consider the fluctuation of the baryon number \footnote{Event-by-event fluctuations of conserved quantities such as the net baryon number can be argued to be a possible signal of the QGP formation and quark-hadron phase transition~\cite{Asakawa:2000wh,Jeon:1999gr}.
Therefore exploring different theoretical aspects relevant to the baryon number fluctuation are important.
Moreover, recent advances in probing the small systems produced in heavy-ion collision experiments also demand a consistent theoretical framework where finite-size effects on various fluctuations can be estimated quantitatively.
To meet such expectations in the present investigation, we consider baryon number fluctuation using spatially smeared or spatially averaged quantum field theory operator to obtain the system size scaling of the fluctuations.}
in the subsystem $S_a$ of the thermodynamic system $S_V$ described by the grand canonical ensemble characterized by the temperature $(T)$ (or its inverse $\beta= 1/T$) and the baryon chemical potential ($\mu_B$).
The volume $V$ of the larger system $S_V$ is larger than the characteristic volume of the subsystem $S_a$. We derive a compact formula that defines quantum fluctuations of the baryon number operator in subsystems of a hot and dense relativistic gas.  Then we apply this formula to get physical insights into situations expected in relativistic heavy-ion collisions. In the calculation  we  use  the metric tensor with the signature $(+1,-1,-1,-1)$.  To denote the scalar product of both four and three-vectors a dot is used , i.e., $a^{\mu}b_{\mu}=a\cdot b= a^0b^0-\vec{a}\cdot\vec{b}$.

The structure of the paper is organized as follows: We start with introducing the basic concepts and definitions in Sec.~\ref{sec:basic}, where we also derive the expression for baryon number density fluctuation. Subsequently, in Sec.~\ref{sec:thermo} we discuss the thermodynamic limit of baryon number density fluctuation. The numerical results of the baryon number density fluctuations are presented in Sec.~\ref{sec:numer}. Finally, in Sec.~\ref{sec:conc} we
interpret and summarize our results.
%%%%%%%%%
%%%%%%%%%%%
\section{Basic concepts and definitions}
\label{sec:basic}
%%%%%%%%%%%%%%%%%%%
We consider a system of spin-$\nicefrac{1}{2}$ particles described by the quantum Dirac field in thermal equilibrium. The field operator is expressed in the standard way as~\cite{Tinti:2020gyh}
\begin{align}
\psi\left(t,\vec{x}\right)=&\sum_r\int\frac{d\mathcal{K}}{\sqrt{2\omega_{\vec{ k}}}}\Big(\mathcal{U}_r^{\pdagger}(\vec{k})a_r^{\pdagger}(\vec{k})e^{-i k \cdot x} +\mathcal{V}_r^{\pdagger}(\vec{k})b_r^{\dagger}(\vec{k})e^{i k \cdot x} \Big),
\label{equ1ver1}
\end{align}
where we use the notation $d\mathcal{K} \equiv {d^3k}/{(2\pi)^3}$, while $a_r^{\pdagger}(\vec{k})$ is the annihilation operator for particles and $b_r^{\dagger}(\vec{k})$ is creation operator for antiparticles. The polarization degree of freedom is represented by the index $r$. The fermionic operators $a_r^{\pdagger}(\vec{k})$ and $b_r^{\dagger}(\vec{k})$ satisfy the canonical anticommutation relations, $\{a_r^{\pdagger}(\vec{k}),a_s^{\dagger}(\vec{k}^{\prime})\} =(2\pi)^3\delta_{rs} \delta^{(3)}(\vec{k}-\vec{k}^{\prime})$ and $ \{b_r^{\pdagger}(\vec{k}),b_s^{\dagger}(\vec{k}^{\prime})\} =(2\pi)^3\delta_{rs}\delta^{(3)}(\vec{k}-\vec{k}^{\prime})$. All the other operators anticommute with each other. The Dirac spinors $\mathcal{U}_r^{\pdagger}(\vec{k})$ and $\mathcal{V}_r^{\pdagger}(\vec{k})$ have normalization $\bar {\mathcal{U}}_r^{\pdagger}(\vec{k}) \mathcal{U}_s^{\pdagger}(\vec{k}) = 2 m \delta_{rs}$ and $\bar{\mathcal{ V}}_r^{\pdagger}(\vec{k}) \mathcal{V}_s^{\pdagger}(\vec{k}) = -2 m \delta_{rs}$, and the quantity $\omega_{\vec{k}}=\sqrt{\vec{k}^2+m^2}$ is the energy of a particle. 

To perform thermal averaging of quantum operators, it is sufficient to know the thermal expectation values of the products of two and four creation and/or annihilation operators (for both particles and antiparticles)~\cite{CohenTannoudji:422962,Itzykson:1980rh,Evans:1996bha}
\begin{align}
& \langle a_r^{\dagger}({\vec{k}})a_s^{\pdagger}({\vec{k}}^{\prime})\rangle=(2\pi)^3\delta_{rs}\delta^{(3)}({\vec{k}}-{\vec{k}}^{\prime})f(\omega_{\vec{k}}),\label{equ2ver1}\\
& \langle a^{\dagger}_r(\vec{k})a^{\dagger}_s(\vec{k}^{\prime})a_{r^{\prime}}^{\pdagger}(\vec{p})a_{s^{\prime}}^{\pdagger}(\vec{p}^{\prime})\rangle\nonumber\\
& =(2\pi)^6 \Big(\delta_{rs^{\prime}}\delta_{r^{\prime}s}\delta^{(3)}(\vec{k}-\vec{p}^{\prime})~\delta^{(3)}(\vec{k}^{\prime}-\vec{p})\nonumber\\
& ~~~~~~~~~~~~~~~~-\delta_{rr^{\prime}}\delta_{ss^{\prime}}\delta^{(3)}({\vec{k}}-\vec{p})~\delta^{(3)}({\vec{k}}^{\prime}-\vec{p}^{\prime})\Big)f(\omega_{{\vec{k}}})f(\omega_{{\vec{k}}^{\prime}}).\label{equ3ver1}
\end{align}
Here $f(\omega_{{\vec{k}}})=1/(\exp(\beta(\omega_{\vec{k}}-\mu_B))+1)$ is the Fermi--Dirac distribution function for particles. For antiparticles, the Fermi--Dirac distribution function differs by the sign of the baryon chemical potential $\mu_B$, i.e. $\bar{f}(\omega_{{\vec{k}}})=1/(\exp(\beta(\omega_{\vec{k}}+\mu_B))+1)$. 
Using the anticommutation relations for $a_r^{\pdagger}({\vec{k}})$, $a_r^{\dagger}({\vec{k}})$, 
$b_r^{\pdagger}({\vec{k}})$, and $b_r^{\dagger}({\vec{k}})$
any other combination of two and four creation and/or annihilation operators can be obtained from Eqs.~\eqref{equ2ver1} and \eqref{equ3ver1}.

Following~Ref.~\cite{Chen:2018cts}, we define the baryon number density operator $\hat{\mathcal{J}}^{0}_a$, associated with the conserved baryon current in a subsystem $S_a$ using a smooth Gaussian profile placed at the origin of the coordinate system, namely
\begin{align}
\hat{\mathcal{J}}^{0}_a = \frac{1}{(a\sqrt{\pi})^3}\int d^3\vec{x}~\hat{\mathcal{J}}^{0}(x)~\exp\left(-\frac{{\vec{x}}^2}{a^2}\right),
\label{equ4ver1}
\end{align}
where $\hat{\mathcal{J}}^{0} = \psi^{\dagger}\psi$. To remove any possible sharp-boundary effects we consider the smooth profile with a length scale $a$ instead of a cubic box~\footnote{Note a similar use of a Gaussian slit by Feynman in~Ref.~\cite{Feynman:100771}}. The thermal expectation value of the normal ordered operator $:\hat{\mathcal{J}}^{0}_a:$ is denoted as $\langle :\hat{\mathcal{J}}^{0}_a :\rangle$. Here standard normal ordering procedure has been introduced to remove an infinite vacuum part coming from zero-point contributions. To determine the baryon number fluctuation of the subsystem $S_a$, we consider the variance
\begin{equation}
\sigma^2(a,m,\beta,\mu_B) = \langle :\hat{\mathcal{J}}^{0}_a: :\hat{\mathcal{J}}^{0}_a: \rangle - \langle :\hat{\mathcal{J}}^{0}_a :\rangle^2\, 
 \label{equ5ver1}
\end{equation}
and the normalized standard deviation as 
\begin{equation} 
\sigma_n(a,m,\beta,\mu_B)= \frac{(\langle:\hat{\mathcal{J}}^{0}_a::\hat{\mathcal{J}}^{0}_a:\rangle- \langle :\hat{\mathcal{J}}^{0}_a :\rangle^2)^{1/2}}{\langle :\hat{\mathcal{J}}^{0}_a :\rangle}.
\label{equ6ver1}
\end{equation}

Using the Fourier mode expansion of the Dirac field operator as given by Eq.~\eqref{equ1ver1}, the normal ordered operator $:\hat{\mathcal{J}}^{0}_a:$ can be obtained as,
\begin{align}
& :\hat{\mathcal{J}}_a^0:= \sum_{r,s}\int \frac{d\mathcal{K}}{\sqrt{2\omega_{\vec{k}}}}\frac{d\mathcal{K}^\prime}{\sqrt{2\omega_{{\vec{k}}^{\prime}}}}\times\nonumber\\
& \bigg[a^{\dagger}_s(\vec{k}^\prime)a^{\pdagger}_r(\vec{k})\bar{\mathcal{U}}^{\pdagger}_s(\vec{k}^\prime)\gamma^0\mathcal{U}_r(\vec{k})~e^{-i(\omega_{\vec{k}}-\omega_{\vec{k}^{\prime}})t}~e^{-\frac{a^2}{4}(\vec{k}-\vec{k}^{\prime})^2}\nonumber\\
&-b^{\dagger}_r(\vec{k})b^{\pdagger}_s(\vec{k}^{\prime})\bar{\mathcal{V}}^{\pdagger}_s(\vec{k}^{\prime})\gamma^0\mathcal{V}_r(\vec{k})~e^{i(\omega_{\vec{k}}-\omega_{\vec{k}^{\prime}})t}~e^{-\frac{a^2}{4}(\vec{k}-\vec{k}^{\prime})^2}\nonumber\\
&+ a^{\dagger}_s(\vec{k}^{\prime})b^{\dagger}_r(\vec{k})\bar{\mathcal{U}}^{\pdagger}_s(\vec{k}^{\prime})\gamma^0\mathcal{V}_r(\vec{k})~e^{i(\omega_{\vec{k}}+\omega_{\vec{k}^{\prime}})t}~e^{-\frac{a^2}{4}(\vec{k}+\vec{k}^{\prime})^2}\nonumber\\
&+ b^{\pdagger}_s(\vec{k}^{\prime})a^{\pdagger}_r(\vec{k})\bar{\mathcal{V}}^{\pdagger}_s(\vec{k}^{\prime})\gamma^0\mathcal{U}_r(\vec{k})~e^{-i(\omega_{\vec{k}}+\omega_{\vec{k}^{\prime}})t}~e^{-\frac{a^2}{4}(\vec{k}+\vec{k}^{\prime})^2}\bigg].
\label{equ7ver1}
\end{align} 
Using the thermal averaging of two creation and/or annihilation operators as given by Eq.~\eqref{equ2ver1}, the thermal expectation value of $:\hat{\mathcal{J}}_a^0:$ has the form
\begin{align}
    \langle :\hat{\mathcal{J}}_a^0: \rangle = 2\int d\mathcal{K} \Big[f(\omega_{\vec{k}})-\bar{f}(\omega_{\vec{k}})\Big]. 
    \label{equ8ver1}
\end{align}
This expression agrees with the standard kinetic-theory definition, with the factor of 2 accounting for the spin degeneracy.

In the next step, using the thermal averaging of four creation and/or annihilation operators as given by Eq.~\eqref{equ3ver1} we obtain
\begin{align}
    & \sigma^2(a,m,\beta,\mu_B)=\langle :\hat{\mathcal{J}}_a^0: :\hat{\mathcal{J}}_a^0:\rangle-\langle :\hat{\mathcal{J}}_a^0: \rangle^2
    \nonumber\\
    &=\int \frac{d\mathcal{K}}{\omega_{\vec{k}}}\frac{d\mathcal{K}^\prime}{\omega_{\vec{k}^{\prime}}}
    (\omega_{\vec{k}}\omega_{\vec{k}^{\prime}}+\vec{k}\cdot\vec{k}^{\prime}+m^2) e^{-\frac{a^2}{2}(\vec{k}-\vec{k}^{\prime})^2}\times\nonumber\\
    & ~~~~~~~~~~~~~~~~~~~~~~~~~~~~~~~~~~~~~~\left[f(\omega_{\vec{k}})\left(1-f(\omega_{\vec{k}^{\prime}})\right)+\bar{f}(\omega_{\vec{k}})(1-\bar{f}(\omega_{\vec{k}^{\prime}}))\right]\nonumber\\
    & -\int \frac{d\mathcal{K}}{\omega_{\vec{k}}}\frac{d\mathcal{K}^\prime}{\omega_{\vec{k}^{\prime}}}
    (\omega_{\vec{k}}\omega_{\vec{k}^{\prime}}+\vec{k}\cdot\vec{k}^{\prime}-m^2) e^{-\frac{a^2}{2}(\vec{k}+\vec{k}^{\prime})^2}\times\nonumber\\
    & ~~~~~~~~~~~~~~~~~~~~~~~~~~~~~~~~~~~~~~~\left[f(\omega_{\vec{k}})(1-\bar{f}(\omega_{\vec{k}^{\prime}}))+\bar{f}(\omega_{\vec{k}})(1-f(\omega_{\vec{k}^{\prime}}))\right].
    \label{equ9ver1}
\end{align}
Although normal ordering removes unwanted vacuum contribution to the baryon number density operator, it is not sufficient to remove all the vacuum divergences in all composite operators. Therefore following Refs.~\cite{Das:2020ddr,Das:2021aar} we discard a divergent temperature and chemical potential independent vacuum term that originally appears in Eq.~\eqref{equ9ver1}. Equation~\eqref{equ9ver1} is the main result which represents a fluctuation of the baryon number density in the subsystem $S_a$. Note that in Eqs.~\eqref{equ8ver1} and \eqref{equ9ver1}, the spin and particle-antiparticle degrees of freedom are included. If we take into account the degeneracy factors $(g)$ due to other internal degrees of freedom then we should do the following replacements: $\langle :\hat{\mathcal{J}}_a^0: \rangle\rightarrow g \langle :\hat{\mathcal{J}}_a^0: \rangle$ and $\sigma^2\rightarrow g \sigma^2$~\cite{Das:2020ddr,Das:2021aar}.  
%%%%%%%%%%%%%%%%%%%
%%%%%%%%%%
%%%%%%%%%%%%%%%%%%%
\begin{figure}[]
\centering
	\includegraphics[scale=0.45]{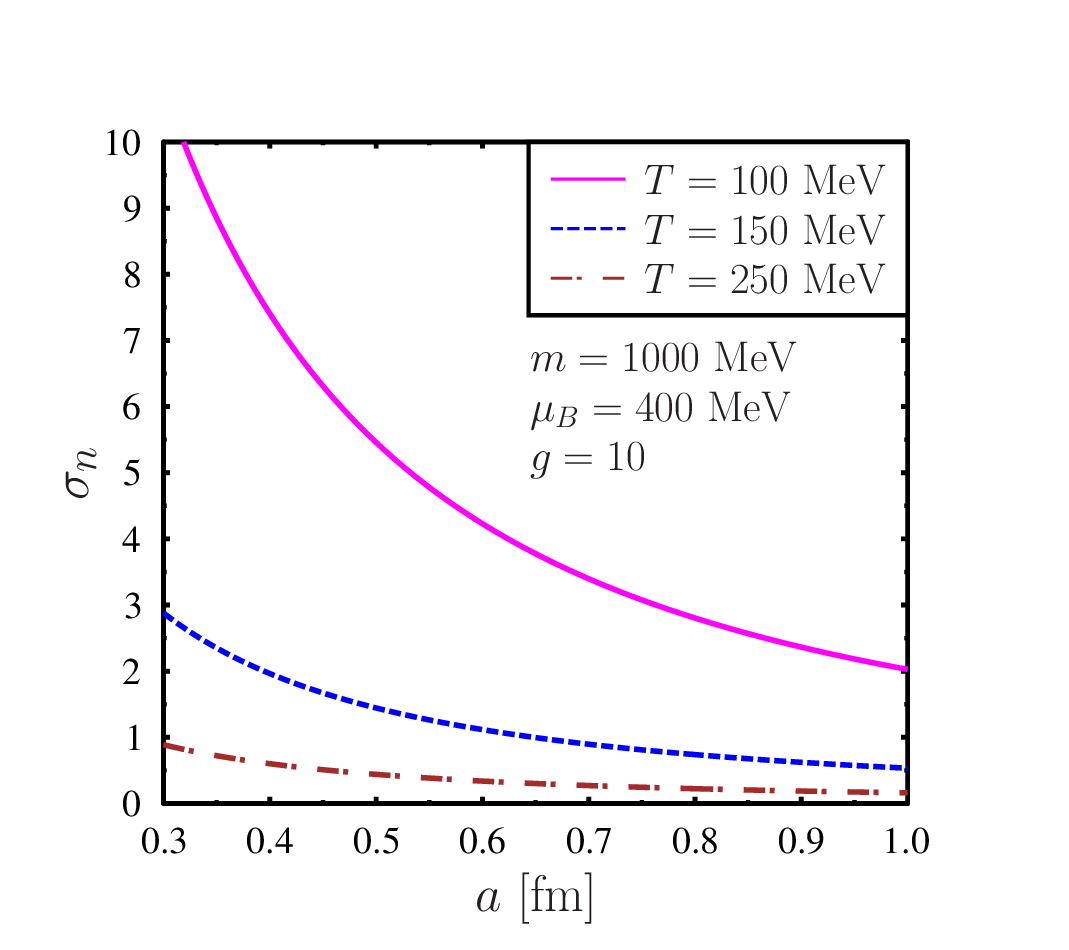}
	\caption{Variation of normalized fluctuation $\sigma_n$ in the subsystem $S_a$ with the scale $a$ for different values of the temperature $T$ and fixed particle mass $m=1000$ MeV and baryon chemical potential $\mu_B=400$ MeV. One may observe that with increasing temperature the normalized fluctuation $\sigma_n$ decreases. }
	\label{fig:1}
\end{figure}
%%%%%%%%%%%%%%%%%%%
\begin{figure}[]
\centering
	\includegraphics[scale=0.45]{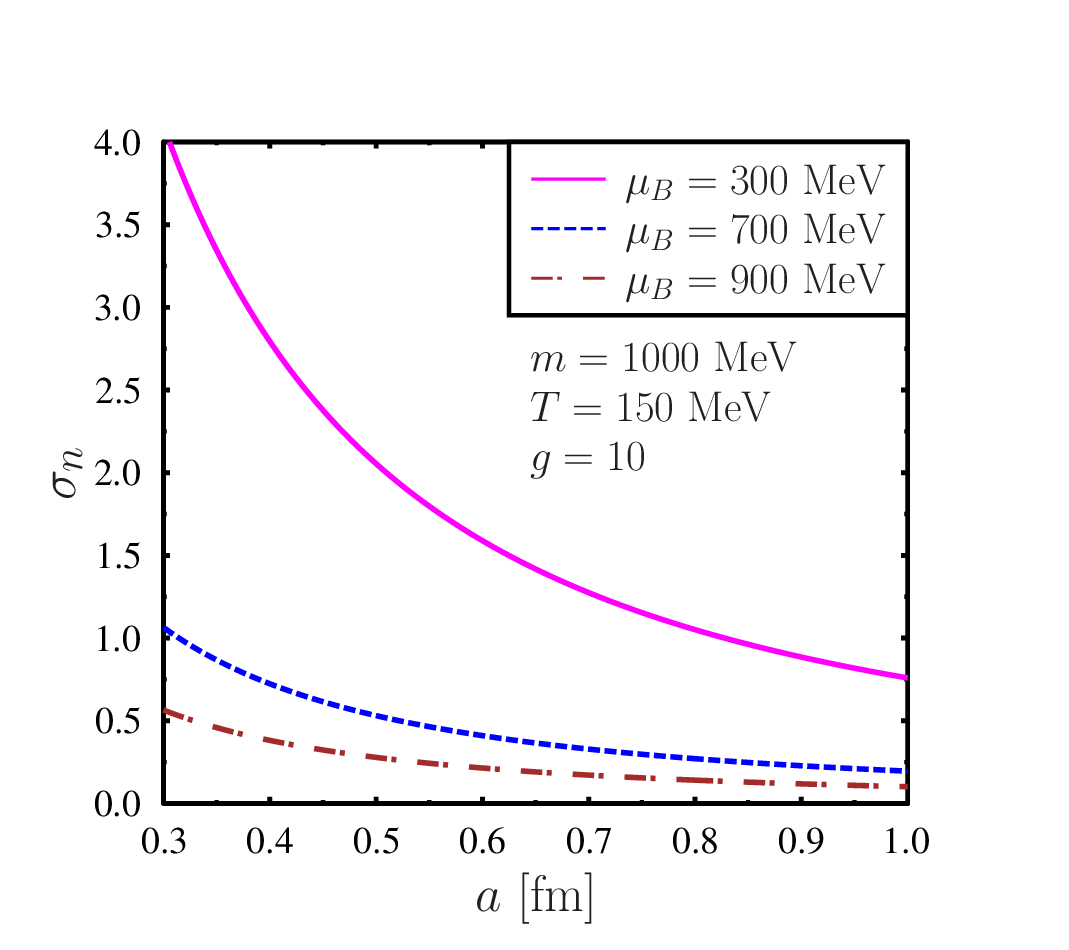}
	\caption{Variation of normalized fluctuation $\sigma_n$ in the subsystem $S_a$ with the scale $a$ for different values of the baryon chemical potential $\mu_B$ and fixed particle mass $m=1000$ MeV and temperature $T=150$ MeV. With an increase in chemical potential the normalized fluctuation $\sigma_n$ decreases.}
	\label{fig:2}
\end{figure}
%%%%%%%%%%%%%%%%%%%
\begin{figure}[]
\centering
	\includegraphics[scale=0.45]{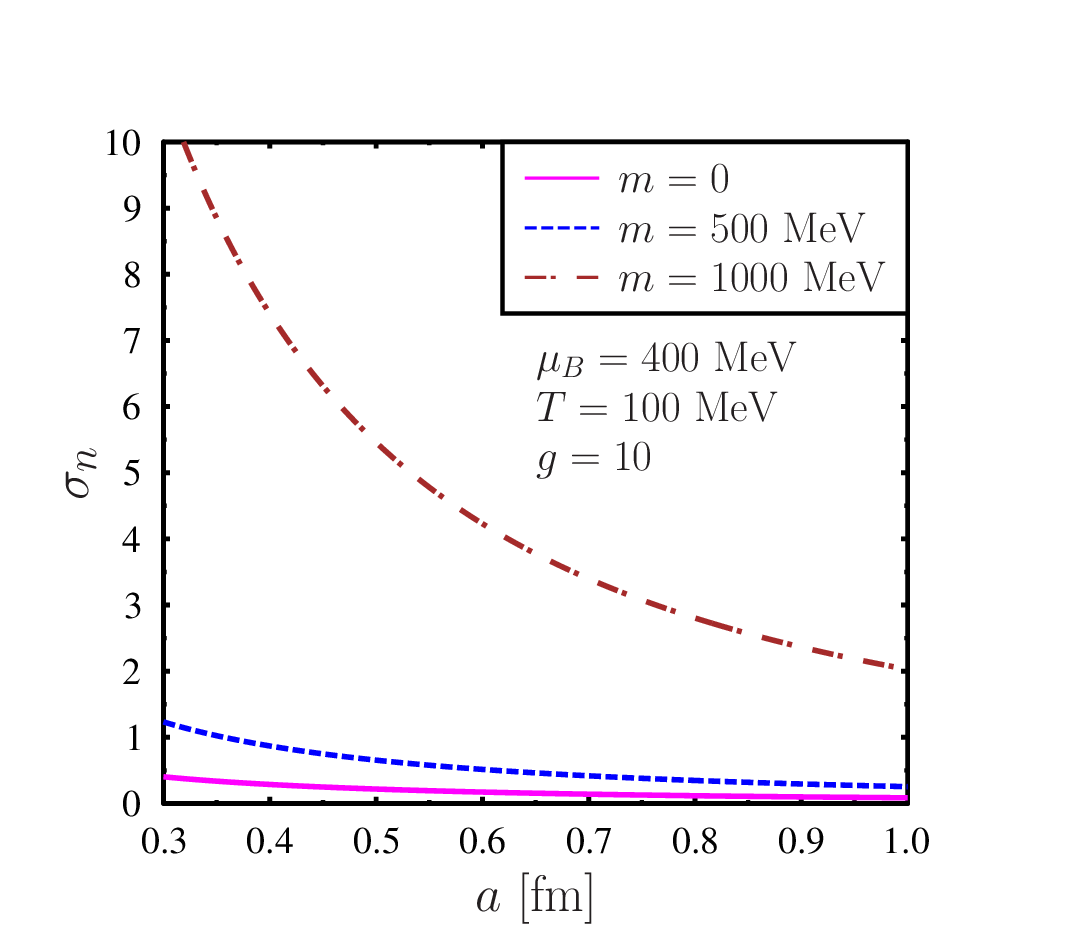}
	\caption{Variation of normalized fluctuation $\sigma_n$ in the subsystem $S_a$ with the scale $a$ for different values of the particle mass and fixed temperature $T = 100$ MeV and baryon chemical potential $\mu_B=400$ MeV. From this figure it is clear that $\sigma_n$ increases with mass of the particle.}
	\label{fig:3}
\end{figure}
%%%%%%%%%%%%%%%%%%%
%%%%%%%%%%
\section{Thermodynamic limit}
\label{sec:thermo}
%%%%%%%%%%%%%%%%%%%
Before we turn to a discussion of our numerical results based on Eq.~\eqref{equ9ver1}, it is important to analyze the thermodynamic limit of $\sigma^2(a,m,\beta,\mu_B)$. Since $S_a$ is a subsystem of the system $S_V$, the thermodynamic limit can be achieved by considering the $a\rightarrow \infty$ limit (still with $a^3 \ll V$). In the thermodynamic limit quantum fluctuation as obtained here should reduce to the expression for the classical statistical fluctuation. To obtain the classical limit of the fluctuations we look into the susceptibilities describing fluctuations in the baryon number obtained for thermal and chemical equilibrium~\cite{Nahrgang:2014fza}
\begin{align}
    \chi_l^{(B)}=\frac{\partial^l(P/T^4)}{\partial(\mu_B/T)^l}\bigg\vert_T.
    \label{equ10ver1}
\end{align}
Here $P$ denotes the thermodynamic pressure. Susceptibilities can also be related to the cumulants of the distribution of baryons, e.g.,
\begin{align}
    & \chi_1^{(B)}=\frac{1}{VT^3}\langle \mathcal{N}_B\rangle=\frac{1}{T^3}\frac{\langle \mathcal{N}_B\rangle}{V}=\frac{n_B}{T^3},\nonumber\\
    & \chi_2^{(B)}=\frac{1}{VT^3}\langle (\triangle \mathcal{N}_B)^2\rangle=\frac{1}{VT^3}\langle (\mathcal{N}_B-\langle \mathcal{N}_B\rangle)^2 \rangle.
    \label{equ11ver1}
\end{align}
Here $n_B\equiv \mathcal{N}_B/V$ is the net baryon number density. Equation~\eqref{equ11ver1} also implies that
\begin{align}
    V \langle (n_B-\langle n_B\rangle)^2 \rangle = T^3\chi_2^{(B)}.
    \label{equ12ver1}
\end{align}
Using Eq.~\eqref{equ10ver1}, the susceptibility $\chi_2^{(B)}$ can also be obtained by taking the appropriate derivative of the thermodynamic pressure. The thermodynamic pressure at finite temperature and baryon chemical potential can be expressed as~\cite{Nahrgang:2014fza}
\begin{align}
    \frac{P}{T^4} = & \frac{2g}{T^3}\int d\mathcal{K}\Bigg[ \ln\bigg(1+\exp\left(-\frac{\omega_{\vec{k}}-\mu_B}{T}\right)\bigg)+\ln\bigg(1+\exp\left(-\frac{\omega_{\vec{k}}+\mu_B}{T}\right)\bigg)\Bigg]. 
    \label{equ13ver1}
\end{align}
Using Eqs.~\eqref{equ10ver1} and \eqref{equ13ver1} we can easily find 
\begin{align}
  \chi_1^{(B)} &=  \frac{\partial(P/T^4)}{\partial(\mu_B/T)}\bigg\vert_T\nonumber\\
  & = \frac{2g}{T^3}\int d\mathcal{K} \Big[f(\omega_{\vec{k}})-\bar{f}(\omega_{\vec{k}})\Big]\nonumber\\
  & = \frac{n_B}{T^3}
  \label{equ14ver1}
\end{align}
and
\begin{align}
    \chi_2^{(B)} &=  \frac{\partial^2(P/T^4)}{\partial(\mu_B/T)^2}\bigg\vert_T\nonumber\\
    & = \frac{2g}{T^3} \int d\mathcal{K} \Big[f(\omega_{\vec{k}})(1-f(\omega_{\vec{k}}))+\bar{f}(\omega_{\vec{k}})(1-\bar{f}(\omega_{\vec{k}}))\Big].
    \label{equ15ver1}
\end{align}
Therefore, we can write
\begin{align}
    T^3\chi_2^{(B)}&=V \langle (n_B-\langle n_B\rangle)^2 \rangle,\nonumber\\
    & = 2g\int d\mathcal{K} \Big[f(\omega_{\vec{k}})(1-f(\omega_{\vec{k}}))+\bar{f}(\omega_{\vec{k}})(1-\bar{f}(\omega_{\vec{k}}))\Big].
    \label{equ16ver1}
\end{align}
In the large volume limit Eq.~\eqref{equ9ver1} should be consistent with Eq.~\eqref{equ16ver1}. This can be verified using the Gaussian representation of the three-dimensional Dirac delta function
\begin{align}
    \delta^{(3)}({\vec{k}}-{\vec{k}^\prime})=\lim_{a \to\infty} \frac{a^3}{(2\pi)^{3/2}}e^{-\frac{a^2}{2}({\vec{k}}-{\vec{k}^\prime})^2}.
    \label{equ17ver1}
\end{align}
This leads us to the following expression
\begin{align}
    &  \lim_{a\to\infty} a^3(2\pi)^{3/2}\bigg[\langle :\hat{\mathcal{J}}_a^0: :\hat{\mathcal{J}}_a^0:\rangle-\langle :\hat{\mathcal{J}}_a^0: \rangle^2\bigg]\nonumber\\
    & =\int \frac{d^3k}{(2\pi)^3} d^3k^{\prime}\frac{1}{\omega_{\vec{k}}}\frac{1}{\omega_{\vec{k}^{\prime}}}
    (\omega_{\vec{k}}\omega_{\vec{k}^{\prime}}+\vec{k}\cdot\vec{k}^{\prime}+m^2)\times \nonumber\\
    &~~~~~~~~~~~~~~~~~~~~~ \left[f(\omega_{\vec{k}})(1-f(\omega_{\vec{k}^{\prime}}))+\bar{f}(\omega_{\vec{k}})(1-\bar{f}(\omega_{\vec{k}^{\prime}}))\right]
    \delta^{(3)}(\vec{k}-\vec{k}^{\prime})\nonumber\\
    & -\int \frac{d^3k}{(2\pi)^3} d^3k^{\prime}\frac{1}{\omega_{\vec{k}}}\frac{1}{\omega_{\vec{k}^{\prime}}}
    (\omega_{\vec{k}}\omega_{\vec{k}^{\prime}}+\vec{k}\cdot\vec{k}^{\prime}-m^2)\times \nonumber\\
    & ~~~~~~~~~~~~~~~~~~~~~~\left[f(\omega_{\vec{k}})(1-\bar{f}(\omega_{\vec{k}^{\prime}}))+\bar{f}(\omega_{\vec{k}})(1-f(\omega_{\vec{k}^{\prime}}))\right]
    \delta^{(3)}(\vec{k}+\vec{k}^{\prime})\nonumber\\
    & = 2\int d\mathcal{K} \Big[f(\omega_{\vec{k}})(1-f(\omega_{\vec{k}}))+\bar{f}(\omega_{\vec{k}})(1-\bar{f}(\omega_{\vec{k}}))\Big],
    \label{equ18ver1}
\end{align}
hence,
\begin{align}
    &  \lim_{a\to\infty} V_a \bigg[\langle :\hat{\mathcal{J}}_a^0: :\hat{\mathcal{J}}_a^0:\rangle-\langle :\hat{\mathcal{J}}_a^0: \rangle^2\bigg]=T^3\chi_2^{(B)},
    \label{equ19ver1}
\end{align}
where $V_a = a^3 (2\pi)^{3/2}$ may be identified as the volume of the ``Gaussian'' subsystem $S_a$. Here we emphasize that the volume scaled fluctuation $V_a\sigma^2$, where $\sigma^2$ is given in Eq.~\eqref{equ9ver1}, and the volume scaled baryon number fluctuation as given in Eq.~\eqref{equ16ver1} differs due to the quantum nature of fluctuation for small system sizes. In the large volume limit, quantum effects are not significant and the volume scaled fluctuation $V_a\sigma^2$ reproduces the standard statistical fluctuation $T^3\chi_2^{(B)}$.

\begin{figure}[]
\centering
	\includegraphics[scale=0.45]{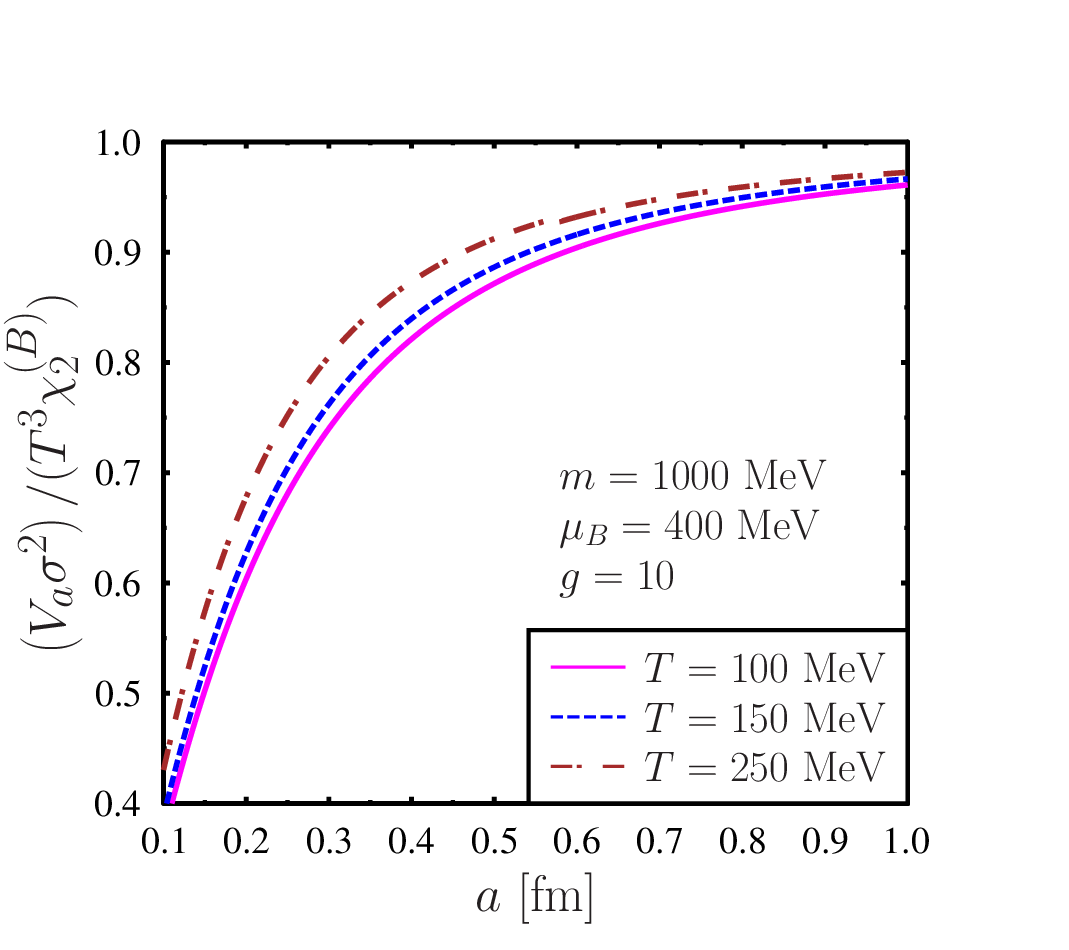}
	\caption{Variation of normalized fluctuation $V_a\sigma^2/(T^3\chi_2^{(B)})$ for different values of temperature ($T$) but with fixed baryon chemical potential ($\mu_B$) and particle mass ($m$).}
	\label{fig:4}
\end{figure}
%%%%%%%%%%%%%%%%%%%
\begin{figure}[]
\centering
	\includegraphics[scale=0.45]{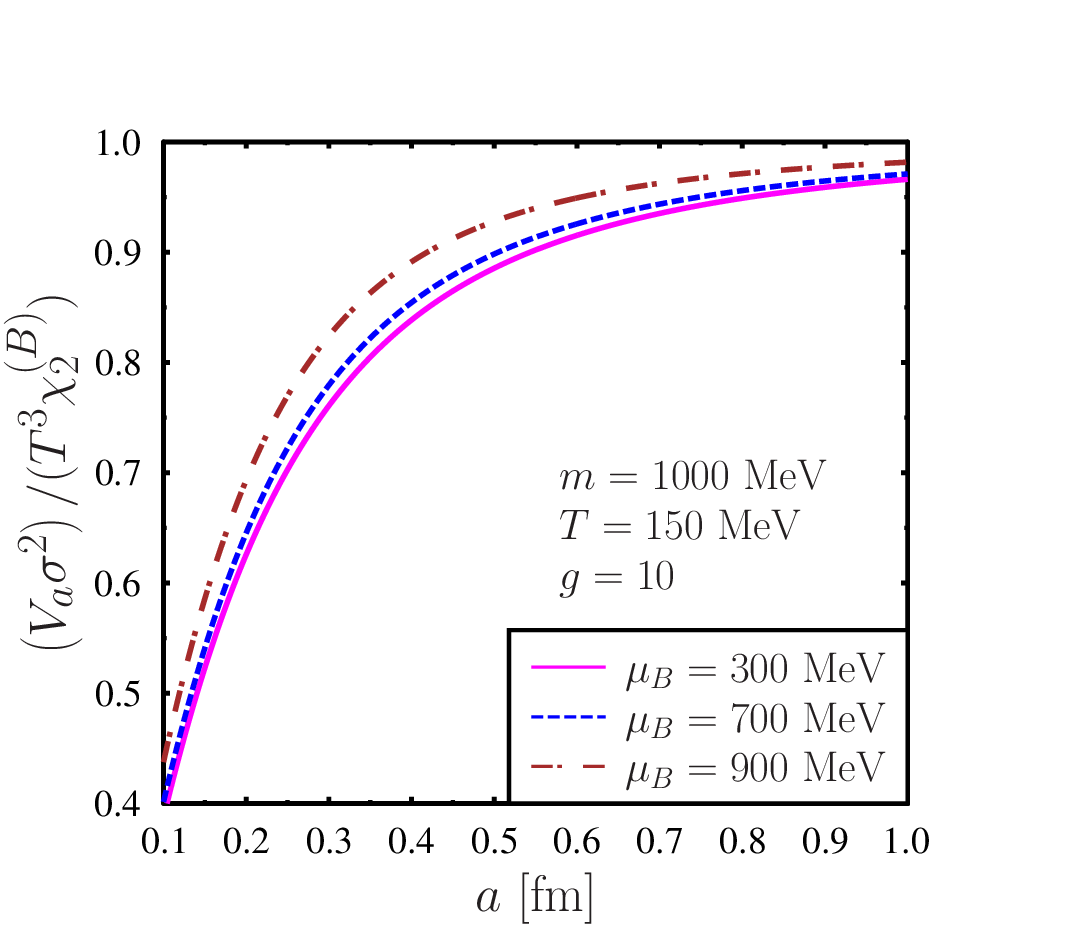}
	\caption{Variation of normalized fluctuation $V_a\sigma^2/(T^3\chi_2^{(B)})$ for different values of baryon chemical potential ($\mu_B$) but with fixed temperature ($T$) and particle mass ($m$).}
	\label{fig:5}
\end{figure}
%%%%%%%%%%%%%%%%%%%
\begin{figure}[]
\centering
	\includegraphics[scale=0.45]{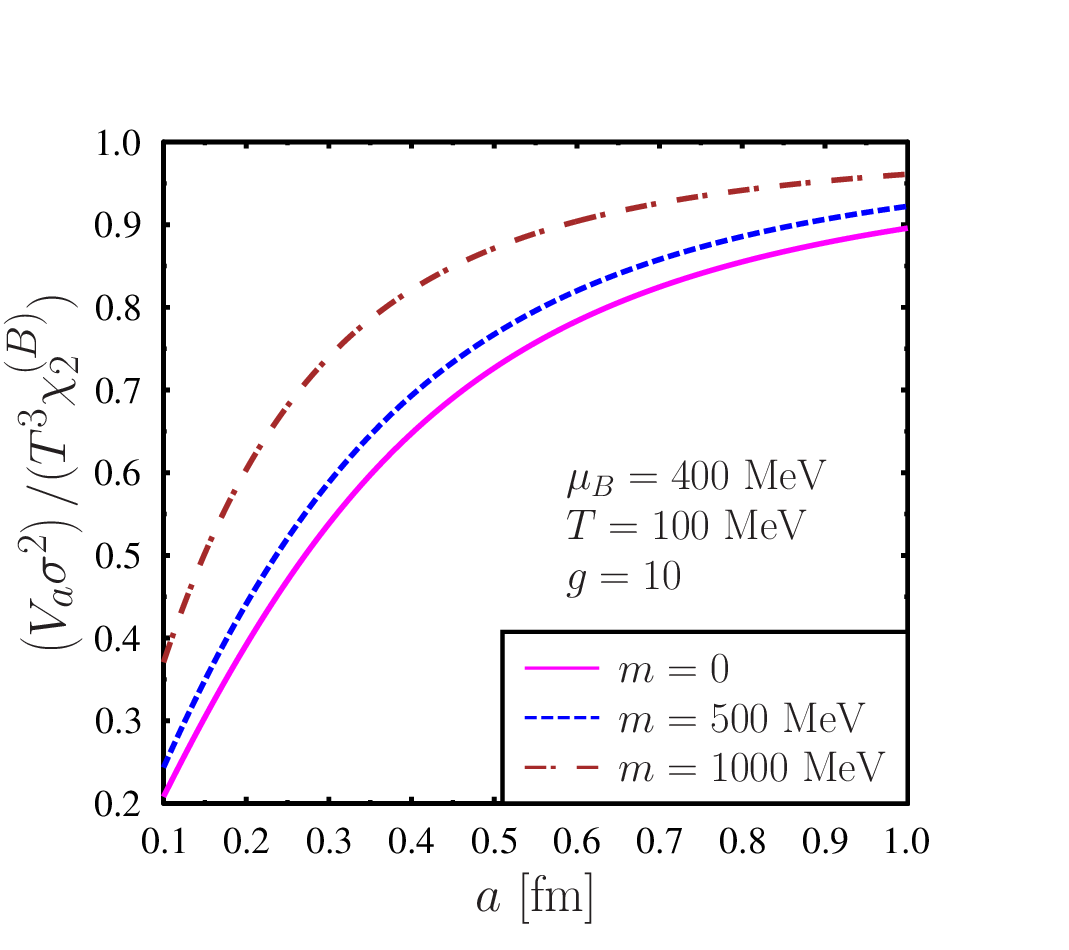}
	\caption{Variation of normalized fluctuation $V_a\sigma^2/(T^3\chi_2^{(B)})$ for different values of particle mass ($m$) but with fixed temperature ($T$) and baryon chemical potential ($\mu_B$)}.
	\label{fig:6}
\end{figure}
%%%%%%%%%%%%%%%%%%%
\section{Numerical results}
\label{sec:numer}
%%%%%%%%%%%%%%%%%%%
Our main result describing fluctuations of the baryon number density in a hot and dense gas of baryons is represented by Eq.~\eqref{equ9ver1}. By straightforward numerical integration, we can obtain the results for any subsystem of size $a$, temperature $T$, baryon chemical potential $\mu_B$, and particle mass $m$. 
In Figs.~\eqref{fig:1}, \eqref{fig:2} and \eqref{fig:3} we show the variation of the normalized fluctuation $\sigma_n$ as defined in Eq.~\eqref{equ6ver1} with the subsystem size $a$ for different values of temperature, baryon chemical potential and mass, respectively.
Keeping in mind the relativistic heavy ion collisions in the regime where large baryon number densities are expected, we consider the following range of the parameters: $100$ MeV $\leq T \leq 250$ MeV, $300$ MeV $\leq \mu_B \leq 900$ MeV and $0$ $\leq m \leq 1000$ MeV. The internal degeneracy factor is taken to be $g= 10$.

Figs.~\eqref{fig:1}, \eqref{fig:2} and \eqref{fig:3} show that the normalized fluctuation $\sigma_n$ decreases with the increase in the size of the subsystem $S_a$, which is the expected behavior of fluctuations. Furthermore, for small system size, the fluctuation is substantially large, which is the manifestation of the quantum mechanical behavior. We note that a priori there is no restriction on $a$, small $a$ means that the system is probed in a very small volume. Due to uncertainty relation, particles in such a small volume have large fluctuations (and consequently large values) of their momenta and energies. This leads to the phenomenon of pair creation and fluctuations observed in our analysis. From our figures, it is clear that the normalized fluctuation $\sigma_n$ decreases with an increase in temperature and baryon chemical potential, however, $\sigma_n$ is large if the particle mass is large. We should mention that variation of $\sigma$ with temperature, baryon chemical potential and mass is opposite to the variation of $\sigma_n$ (not shown here). $\sigma_{n}$ can be considered as a dimensionless measure of quantum fluctuation.
If $\sigma_{n} \lesssim 1$ then one may infer that the quantum effects are small.   

In Figs.~\eqref{fig:4}, \eqref{fig:5} and \eqref{fig:6} we demonstrate the thermodynamic limit of the volume scaled fluctuation $V_a\sigma^2$. From Eq.~\eqref{equ19ver1} we conclude that in the thermodynamic limit, i.e., in the $a\rightarrow \infty$ limit, $V_a\sigma^2/(T^3\chi_2^{(B)})$ should approach unity. This property can be seen clearly in Figs.~\eqref{fig:4}, \eqref{fig:5} and \eqref{fig:6}, where for $a>1$ fm quantum fluctuations approaches the classical limit which also depend upon the temperature ($T$), baryon chemical potential ($\mu_B$) and particle mass ($m$). Furthermore, for small system size the quantum fluctuations can be significant \footnote{In Figs.~\eqref{fig:4}-\eqref{fig:6}, we demonstrate that in the large volume limit the volume scaled quantum statistical fluctuation $(V_a\sigma^2)$ approaches its classical limit $(T^3\chi_2^{(B)})$.
One may observe from Figs.~\eqref{fig:4}-\eqref{fig:6} that with smaller system sizes the deviation of $V_a\sigma^2$ from its classical limit $(T^3\chi_2^{(B)})$ increases.
In other words, the ratio $V_a\sigma^2/(T^3\chi_2^{(B)})$ deviates from unity for small values of `$a$'. Therefore we can conclude that for small system sizes effects of quantum fluctuations can be significant. One should also note that in Figs.~\eqref{fig:4}-\eqref{fig:6} we plotted the ratio $V_a\sigma^2/(T^3\chi_2^{(B)})$. This ratio has a smaller value for the $a\rightarrow 0$ limit, however it does not mean that the quantum fluctuations are small for the $a\rightarrow 0$ limit.}.\\
%%%%%%%%%%%%%%%%%%%
\section{Conclusion}
\label{sec:conc}
%%%%%%%%%%%%%%%
In this work, we have analyzed quantum baryon-number fluctuations in subsystems of a hot and dense relativistic gas of fermions and found that they diverge for small system sizes. On the other hand, our results agree with the results known from statistical physics for sufficiently large system size $a$. In this way, we have delivered a useful formula that accounts for both statistical and quantum features of the fluctuations. The numerical results have been obtained for a broad range of thermodynamic parameters. They may be useful to interpret and shed new light on the heavy-ion experimental data.
In particular, in the context of the search for the QCD critical point our results (after generalization to describe a hadron gas) may serve as a more appropriate reference point in the case of small systems, where the enhanced fluctuation may be a quantum effect discussed here.

We believe that the treatment of the quantum statistical fluctuation of baryon number as given in the present study is very novel and its connection to experiments is of paramount importance. Therefore, our goal for future analysis is to present a detailed study on how such fluctuations for a spatially smeared operator can be probed experimentally using the observed particle spectra.
Note that due to the collective behavior of the strongly interacting plasma the position space fluctuations may be correlated with the momentum space observables. Quantum statistical fluctuations can play an important role in the dynamics of the net-baryon density in low-energy collision experiments.
To incorporate such effects of quantum fluctuations, one may look into the stochastic dynamics of the net-baryon density which become important near the QCD critical endpoint~\cite{PhysRevD.70.056001,Nahrgang:2017hkh}.
Near the critical endpoint, the net-baryon density can exhibit diffusive dynamics, and fluctuations in the net-baryon number are enhanced~\cite{PhysRevD.60.114028,PhysRevLett.81.4816}.
Moreover, non-equilibrium effects can become relevant due to the fast dynamics of the expanding QCD medium.
Fluctuations can also be coupled to the hydrodynamic models to obtain a dynamical evolution which can leave traces of quantum fluctuations on the event-by-event distributions of net-baryon number~\cite{PhysRevC.93.021902,PhysRevC.87.014907}.

%%%%%%%%%%%%%%%%%%%
%%%%%%%%%%%%%%%%%%%

%\begin{acknowledgements}
This research was supported in part by the Polish National Science Centre Grants No. 2016/23/B/ST2/00717 and No. 2018/30/E/ST2/00432.
%\end{acknowledgements}

\bibliography{fluctuationRef.bib}{}
\bibliographystyle{spphys}

\end{document}